\documentclass[sigplan,dvipsnames,screen]{acmart}
\copyrightyear{2025}
\acmYear{2025}
\setcopyright{cc}

\makeatletter
\def\@ACM@copyright@check@cc{}
\makeatother
\setcctype{by}

\acmConference[ASPLOS '25]{Proceedings of the 30th ACM International Conference on Architectural Support for Programming Languages and Operating Systems, Volume 2}{March 30-April 3, 2025}{Rotterdam, Netherlands}
\acmBooktitle{Proceedings of the 30th ACM International Conference on Architectural Support for Programming Languages and Operating Systems, Volume 1 (ASPLOS '25), March 30-April 3, 2025, Rotterdam, Netherlands}\acmDOI{10.1145/3669940.3707230}
\acmISBN{979-8-4007-0698-1/25/03}

\usepackage[colorinlistoftodos]{todonotes}

\usepackage[many]{tcolorbox}

\usepackage{graphicx}
\usepackage{subfig}

\usepackage[export]{adjustbox}
\usepackage{graphicx}
\usepackage{mdframed}
\usepackage{textcomp}
\usepackage{pgfplots}
\usepackage{tikz}
\usepackage{pgf-pie}
\usepackage{algpseudocode}
\usepackage{lipsum}
\usepackage[linesnumbered,titlenumbered,ruled,noend]{algorithm2e}
\usepackage{algorithmicx}
\usepackage{todonotes}
\usepackage{multirow}
\usepackage{fancyhdr}
\usepackage{fixltx2e}
\usepackage[all]{nowidow}
\definecolor{lstgrey}{rgb}{0.95,0.95,0.95}
\usepackage{listings}
\usepackage{textcomp}
\usepackage{svg}
\usepackage{tikzscale}

\usepackage{enumitem}
\usepackage{tabu,multirow,makecell}
\usepackage{pifont}
\renewcommand{\textrightarrow}{$\rightarrow$}

\usepackage{etoolbox}
\usepackage{xcolor}
\usepackage{pdfpages} 
\usepackage{listings}

\definecolor{dkgreen}{rgb}{0,0.6,0}
\definecolor{gray}{rgb}{0.5,0.5,0.5}
\definecolor{mauve}{rgb}{0.58,0,0.82}
\definecolor{OliveGreen}{rgb}{0,0.6,0}
\definecolor{mahogany}{rgb}{0.75, 0.25, 0.0}
\definecolor{darkmidnightblue}{rgb}{0.0, 0.2, 0.4}
\definecolor{navyblue}{rgb}{0.0, 0.0, 0.5}
\definecolor{apricot}{rgb}{0.98, 0.81, 0.69}
\definecolor{antiquewhite}{rgb}{0.98, 0.92, 0.84}
\definecolor{brickred}{rgb}{0.8, 0.25, 0.33}
\definecolor{bananamania}{rgb}{0.98, 0.91, 0.71}
\definecolor{bisque}{rgb}{1.0, 0.89, 0.77}
\definecolor{champagne}{rgb}{0.97, 0.91, 0.81}
\definecolor{eggshell}{rgb}{0.94, 0.92, 0.84}
\definecolor{darkolivegreen}{rgb}{0.33, 0.42, 0.18}
\definecolor{phthalogreen}{rgb}{0.07, 0.21, 0.14}
\definecolor{richblack}{rgb}{0.0, 0.25, 0.25}
\definecolor{anti-flashwhite}{rgb}{0.95, 0.95, 0.96}
\definecolor{aliceblue}{rgb}{0.94, 0.97, 1.0}
\definecolor{azuremist}{rgb}{0.94, 1.0, 1.0}

\usepackage[flushleft]{threeparttable}

\definecolor{pie1}{gray}{0.2}
\definecolor{pie2}{gray}{0.4}
\definecolor{pie3}{gray}{0.6}
\definecolor{pie4}{gray}{0.8}
\definecolor{pie5}{gray}{0.9}

\definecolor{pie11}{gray}{0.1}
\definecolor{pie12}{gray}{0.25}
\definecolor{pie13}{gray}{0.4}
\definecolor{pie14}{gray}{0.55}
\definecolor{pie15}{gray}{0.70}
\definecolor{pie16}{gray}{0.85}

\definecolor{lbcolor}{rgb}{0.9,0.9,0.9}  

\makeatletter
\patchcmd{\@algocf@start}
  {-1.5em}
  {0pt}
  {}{}
\makeatother

\def\BibTeX{{\rm B\kern-.05em{\sc i\kern-.025em b}\kern-.08em
    T\kern-.1667em\lower.7ex\hbox{E}\kern-.125emX}}

\usetikzlibrary{shadows,patterns,shapes,arrows,decorations.pathmorphing,backgrounds,positioning,fit,plotmarks,calc,spy}
\usepgfplotslibrary{patchplots}
\definecolor{darktan}{rgb}{0.57, 0.51, 0.32}
\definecolor{desert}{rgb}{0.76, 0.6, 0.42}
\definecolor{desertsand}{rgb}{0.93, 0.79, 0.69}
\pgfplotsset{compat=1.5,
	/pgfplots/ybar legend/.style={
		/pgfplots/legend image code/.code={%
			\draw[##1,/tikz/.cd,bar width=3pt,yshift=-0.2em,bar shift=0pt]
			plot coordinates {(0cm,0.8em)};},
	},
}

\usetikzlibrary{pgfplots.statistics}
\usetikzlibrary{snakes}
\renewcommand{\arraystretch}{1.5}

\makeatletter
\def\endthebibliography{%
	\def\@noitemerr{\@latex@warning{Empty `thebibliography' environment}}%
	\endlist
}
\makeatother

\usepackage{enumerate}
\usepackage{xurl}

\usepackage[]{hyperref}

\begin{document}

\title{Enhancing CGRA Efficiency Through Aligned Compute and Communication Provisioning}
\author{Zhaoying Li}
\authornote{Both authors contributed equally to this research}
\email{zhaoying@comp.nus.edu.sg}
\affiliation{%
  \institution{National University of Singapore}
  \country{Singapore}
}

\author{Pranav Dangi}
\authornotemark[1]
\email{dangi@comp.nus.edu.sg}
\affiliation{%
  \institution{National University of Singapore}
  \country{Singapore}
}

\author{Chenyang Yin}
\email{ycy@stu.pku.edu.cn}
\affiliation{
  \institution{Peking University}
  \country{China}
}

\author{Thilini Kaushalya Bandara}
\email{thilini@comp.nus.edu.sg}
\affiliation{%
  \institution{National University of Singapore}
  \country{Singapore}
}

\author{Rohan Juneja}
\email{rohan@comp.nus.edu.sg}
\affiliation{%
  \institution{National University of Singapore}
  \country{Singapore}
}

\author{Cheng Tan}
\email{chengtan@google.com}
\affiliation{%
  \institution{Google}
  \country{United States}
}

\author{Zhenyu Bai}
\email{zhenyu.bai@nus.edu.sg}
\authornote{Corresponding Author}
\affiliation{%
  \institution{National University of Singapore}
  \country{Singapore}
}

\author{Tulika Mitra}
\email{tulika@comp.nus.edu.sg}
\affiliation{%
  \institution{National University of Singapore}
  \country{Singapore}
}

\begin{abstract}
Coarse-grained Reconfigurable Arrays (CGRAs) are domain-agnostic accelerators that enhance the energy efficiency of resource-constrained edge devices. The CGRA landscape is diverse, exhibiting trade-offs between performance, efficiency, and architectural specialization. However,  CGRAs often overprovision communication resources relative to their modest computing capabilities.
This occurs because the theoretically provisioned programmability for CGRAs often proves superfluous in practical implementations.

In this paper, we propose \textit{Plaid}, a novel CGRA architecture and compiler that aligns compute and communication capabilities, thereby significantly improving energy and area efficiency while preserving its generality and performance.
We demonstrate that the dataflow graph, representing the target application,  can be decomposed into smaller, recurring communication patterns called \textit{motifs}.
The primary contribution is the identification of these structural motifs within the dataflow graphs and the development of an efficient collective execution and routing strategy tailored to these motifs.
The \textit{Plaid} architecture employs a novel collective processing unit that can execute multiple operations of a motif and route related data dependencies together.
The \textit{Plaid} compiler can hierarchically map the dataflow graph and judiciously schedule the motifs. 
Our design achieves a 43\% reduction in power consumption and 46\% area savings compared to the baseline high-performance spatio-temporal CGRA, all while preserving its generality and performance levels. In comparison to the baseline energy-efficient spatial CGRA, \textit{Plaid} offers a 1.4$\times$ performance improvement and a 48\% area savings, with almost the same power.
\end{abstract}



\maketitle 
\pagestyle{plain} 

\section{Introduction}
The surge in new applications like Machine Learning has led to the rapid development of accelerators tailored to specific domains or frequently used computational kernels~\cite{dally2020domain}. Although these specific accelerators excel in performance and power efficiency,  edge devices cannot accommodate or utilize many accelerators due to strict power and area constraints~\cite{hadi2011dark, wall2019fuchs}. Coarse-grained Reconfigurable Arrays (CGRAs)~\cite{li2022coarse, sustainable-cgras} offer a balanced solution for performance, efficiency, and programmability, making them ideal for edge acceleration.

Several academic~\cite{mei2003adres, karunaratne2017hycube, nowatzki2017stream, weng2020hybrid, torng2021ultra, tan2021arena, tan2021aurora,dadu2021polygraph,gobieski2022riptide,serafin2023pipestitch, melchert2023apex,feng2023amber,tan2023vecpac, voitsechov2018inter, li2021chordmap, li2022lisa, tan2024micro, wu2023flip, wijerathne2019cascade, wijerathne2021himap, wijerathne2022panorama, wijerathne2022morpher, wijerathne2023accelerating} and commercial ~\cite{kim2014ulp, fleming2020processors, fujii2018drp, prabhakar2017plasticine} CGRA architectures have been proposed over the years. A typical CGRA consists of an array of Processing Elements (PEs) connected by an on-chip network (NoC).  Each PE comprises a computing unit (typically an ALU), registers to store the temporal data, a router for interconnections, and a configuration memory to store the instructions statically generated by the compiler for the above hardware modules. The compiler maps the target application, represented by a Dataflow Graph (DFG), onto the CGRA. Computing units execute the DFG nodes, while routers and registers handle the data dependency, i.e., the edges representing communication among these nodes.

The existing landscape of CGRAs navigates trade-offs between performance, efficiency, and generality, where generality refers to the versatility of the scope of target applications or kernels that can be mapped onto the architecture. As shown in Table~\ref{tab:cgra_comparison}, existing CGRAs fall into three main categories: spatio-temporal CGRAs, spatial CGRAs, and specialized CGRAs.

\begin{table}[]
\centering
\resizebox{\linewidth}{!}{%
\begin{tabular}{|l|l|l|l|}
\hline
\textbf{CGRA Architecture} & \textbf{Performance} & \textbf{Energy Efficiency} & \textbf{Generality} \\ \hline
\textbf{Spatio-temporal} & \multirow{3}{*}{High}& \multirow{3}{*}{Low}& \multirow{3}{*}{High}\\
 UE-CGRA~\cite{torng2021ultra}, HyCUBE~\cite{karunaratne2017hycube}& & &\\
ADRES~\cite{mei2003adres}, MorphoSys~\cite{singh2000morphosys}& & & \\ \hline
\textbf{Spatial } & \multirow{2}{*}
{Medium* or High}& \multirow{2}{*}
{High}& \multirow{2}{*}
{Medium}\\
 SNAFU~\cite{gobieski2021snafu}, Riptide~\cite{ gobieski2022riptide}& & &\\ \hline
\textbf{Specialized } & \multirow{3}{*}{High or Ultra-High}& \multirow{3}{*}{High}& \multirow{3}{*}{Low}\\
REVAMP~\cite{revamp}, REVEL~\cite{weng2020hybrid}&  & & \\
 VecPac~\cite{tan2023vecpac}, APEX~\cite{melchert2023apex}& & &\\ \hline
\rowcolor{lightgray} \textbf{Ours (Plaid)} & High & High & High \\ \hline
\end{tabular}
}
\caption*{\scriptsize{* Performance degrades when partitioning complex DFGs to be mapped spatially}}
\caption{Reconfigurable architecture landscape}
\label{tab:cgra_comparison}
\end{table}

A spatio-temporal architecture allows each PE to reconfigure to a new instruction every cycle, offering higher flexibility. This means that the ALU in each PE can execute a new operation and send the data to a new neighbour every cycle. In contrast, a purely spatial architecture relies on spatial dataflow-based mapping for a code segment, maintaining a fixed configuration of compute and communication during the execution of that segment. 
Theoretically, a higher degree of reconfigurability enhances flexibility at the cost of higher power consumption. On the other hand, fully spatial mapping ensures extremely low power consumption but with possibly lower performance. 
Specialized CGRAs, in contrast to the prior ones, are optimized or tuned for a particular application suite or domain. These optimizations are only suitable for a relatively narrow scope of applications and can render the CGRA to be less efficient for others, thereby reducing their versatility. An ideal CGRA would achieve good performance and efficiency while still preserving generality.

Existing works intrinsically facilitate effective execution at the granularity of a single DFG node on a PE. However, they lack insight into the alignment between communication and computation for the collective execution of the entire DFG. 
First, each PE pairs the ALU with a router designed with adequate degrees of freedom to allow each ALU to individually communicate with any other ALU around it through the NoC. Such highly powerful routers help PEs handle peak routing congestion. An unforeseen consequence is that total communication resources are largely overprovisioned compared to the compute capabilities of the CGRA. This is because such congestion is only faced intermittently throughout total execution. 
Second, as each PE typically executes one node at a time, the inherent individuality in execution renders the programmability less efficient. For example,  even simple inter-PE communication, corresponding to a single edge between two DFG nodes, necessitates configuring at least two routers. This bloats the configuration memory, contributing to 48\% of the overall power consumption as shown in Figure~\ref{fig:intro_power}(a). With the aforementioned resource over-provisioning, the programmability becomes extremely costly.

\begin{figure}[t]
    \centering
    \includegraphics[width=\linewidth, page=14]{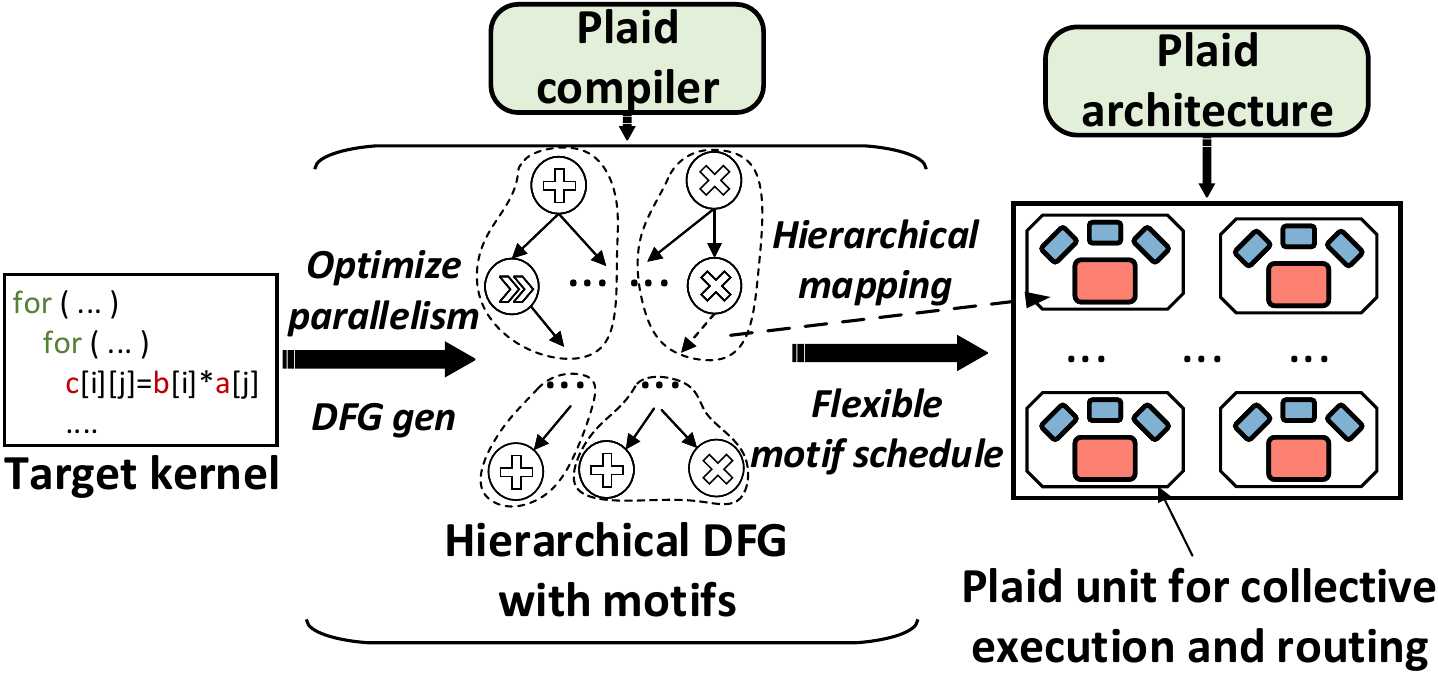}
    \caption{Overview of Plaid co-designed architecture and compiler}
    \label{fig:overview}
\end{figure}

A straightforward approach to the realignment would be trimming communication resources to a bare minimum. While this strategy is conceptually simple, it can hinder reconfigurability in the architecture and degrade the performance.
Therefore, a more nuanced solution should hypothetically involve redesigning the fundamental blocks of the CGRA architecture and the compiler. This redesign should collectively align compute and communication provisioning and improve efficiency while maintaining programmability,  considering the inherent characteristics of DFGs and hardware. 
\newline

\noindent\textbf{Goal and Approach:} In this work, we propose a next-gen CGRA architecture and compiler, named \textit{Plaid},  with a hierarchical execution paradigm to address the aforementioned limitation. \textit{Plaid}\footnote{  Despite the fundamental change in CGRA design, \textit{Plaid} is still a spatio-temporal CGRA. When we refer to spatio-temporal CGRA in this paper, we refer to the typical design of spatio-temporal CGRAs.} aims to significantly reduce power and area while maintaining the performance and generality of current spatio-temporal CGRAs. 
Figure~\ref{fig:overview} demonstrates the end-to-end framework of \textit{Plaid} that enables this alignment in computation and communication. \textit{Plaid} takes a target kernel as an annotated C code as an input, converts it into a hierarchical DFG with recurring patterns of communication, i.e., motifs, and maps them onto the proposed CGRA architecture.
Below, we delineate the proposed design and highlight the innovative features of our approach:

\begin{figure}[t]
\subfloat[Spatio-temporal]{ \begin{tikzpicture}[scale=0.35, align=left]
    \tikzstyle{every node}=[font=\fontsize{6pt}
    {6pt}\selectfont]

       \pie[pos = {0,0}, explode=0.2,  radius=3.5, text = legend,  font={\fontsize{8pt}{8pt}\selectfont}, color = {pie1, pie2, pie3, pie5, pie4}]  
     {15/{Routers}, 29/{Communication\\config},  19/{Compute\\config}, 28/{Compute}, 9/{others}}

 \end{tikzpicture}
 }
\subfloat[Plaid]{ \begin{tikzpicture}[scale=0.35, align=left]]
    \tikzstyle{every node}=[font=\fontsize{6pt}
    {6pt}\selectfont]
    
    
     \node[anchor=west] at (-4.5, 4) {\small 43\% power reduction $\downarrow$};

    \pie[radius = 3.43, explode=0.2, font={\fontsize{10pt}{10pt}\selectfont}, color = {pie1, pie2, pie3, pie5, pie4}]  
    {8/{}, 16/{},  14/{}, 49/{}, 12/{}}

 \end{tikzpicture}}
    \caption{Power distribution for Plaid and a current state-of-the-art spatio-temporal CGRA. Plaid's execution paradigm reduces CGRA fabric power by 43\% while maintaining performance and generality. }
    \label{fig:intro_power}
\end{figure}
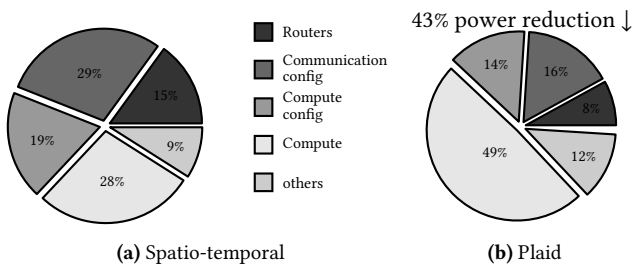

First, we introduce a novel hierarchical execution paradigm that capitalizes on communication patterns inherent in the DFG structure, termed as "motifs." 
These motifs, composed of multiple nodes, demonstrate distinctive yet straightforward internal connections, enabling the CGRA to execute these nodes and route related data dependencies collectively with minimal hardware overhead.

Second, we introduce a novel CGRA architecture \textit{Plaid}, aligning the compute and communication resources for both low-level and high-level execution.
This architecture features a novel hierarchical on-chip network, composed of global and local routers, and re-organizes the compute units for collective resource alignment and motif computing. 


Third, we develop an end-to-end toolchain consisting of a compiler that takes annotated loops in C code as input and maps them onto the \textit{Plaid} architecture written in RTL. The compiler automatically identifies motifs within DFGs, flexibly schedules motifs, and hierarchically maps DFGs onto the CGRA, ensuring excellent generality and performance.

\noindent We make the following contributions:
\begin{itemize}[leftmargin=*]
\item  Identification of a major limitation prevalent in previous CGRAs regarding the alignment of compute and communication resources.
\item Insight into DFG's structural regularity, offering a new dimension to enhance efficiency beyond current execution.
\item The first proposal of a hierarchical on-chip network within a single CGRA, to the best of our knowledge.
\item An effective compiler to map DFG with motifs to achieve hierarchical execution.
\item We demonstrate that we can further improve energy and area efficiency with domain-specific optimization.
\end{itemize}

\noindent\textbf{Results:} Our design reduces 43\% power and saves 46\% area compared to the baseline high-performance spatio-temporal CGRA, while maintaining the generality and performance. At the same time, compared to the baseline energy-efficient spatial CGRA, \textit{Plaid} delivers 1.4$\times$ performance and saves 48\% area while achieving almost the same power. Furthermore, \textit{Plaid} can achieve 1.22$\times$ energy efficiency and 1.25$\times$ area efficiency even compared to a domain-optimized CGRA.

\section{Background And Motivation}

\begin{figure}[t]
    \centering
    \includegraphics[width=0.95\linewidth, page=3]{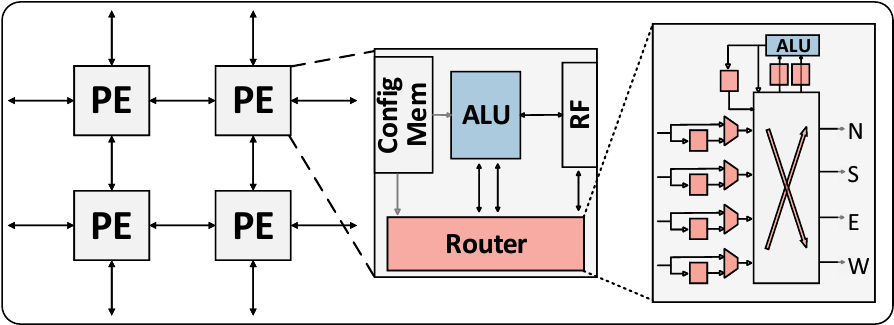}
    \caption{Spatio-temporal CGRA example}
    \label{fig:motivation_cgra}
\end{figure}

In contemporary CGRAs~\cite{balasubramanian2022pathseeker,cong2014fully,dadu2022taskstream,dadu2019towards,das2017142mops,dave2018ureca,hamzeh2014branch,kim2009hierarchical,miyamori1999remarc,nguyen2021fifer,nicol2017coarse,nowatzki2018hybrid,nowatzki2015exploring,ozaki2011cool,parashar2013triggered,park2009polymorphic,sankaralingam2021mozart,vilim2020gorgon,voitsechov2014single,voitsechov2018inter}, the PE is typically centered around an ALU as its core component. A PE usually executes one DFG node at a time and simultaneously routes data to and from neighbouring PEs. Figure~\ref{fig:motivation_cgra} shows a representative CGRA architecture, detailing the internals of a PE, along with a router and corresponding registers for data buffering. Each PE contains a configuration memory to store instructions that reconfigure the ALU and routers every cycle.

\begin{figure}[t]
\centering
\begin{minipage}{.5\linewidth}
\begin{mdframed}[backgroundcolor=aliceblue]
\lstset{escapechar=|} 
\begin{lstlisting}[basicstyle=\linespread{0.9}\footnotesize]
|{\color{black}int main| (){
 |{\color{darkolivegreen} \bf\it \# pragma Plaid}|
 |{\color{navyblue}for | (|{\color{navyblue} i: 0->N|){
  |{\color{navyblue}for | (|{\color{navyblue} j: 0->M|){
   c[i][j] = b[i]*k +a[i]*j;
   k[i] = d[i] >> 4;
   ...
   out[i] += c[i][j] +f[j];
}}}

\end{lstlisting}
\end{mdframed}
\label{source_code}
\end{minipage}
\hfill
\begin{minipage}{0.47\linewidth}
        \includegraphics[width=\linewidth, page=5]{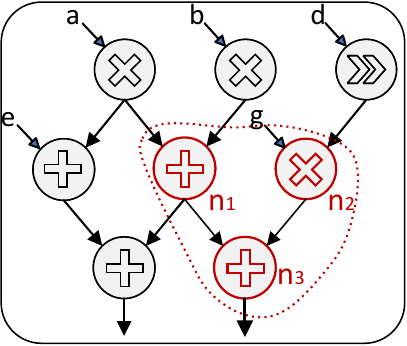}
\end{minipage}
\caption{Example of a DFG corresponding to an annotated C code}
\label{fig:motivation_dfg}
\end{figure}

Figure~\ref{fig:motivation_dfg} showcases an example DFG corresponding to a loop, annotated with a pragma in the C code.
Each DFG node signifies a compute or load-store operation. Typically, each DFG node is mapped to a CGRA PE through a statically compiled instruction configuration. This instruction includes the computation to be executed on the ALU and the communication required with neighbouring PEs. This communication encoding is crucial for selecting the appropriate datapath via a series of multiplexers and crossbars in each cycle.

\begin{figure}[t]
    \centering
    \includegraphics[width=0.8\linewidth, page=6]{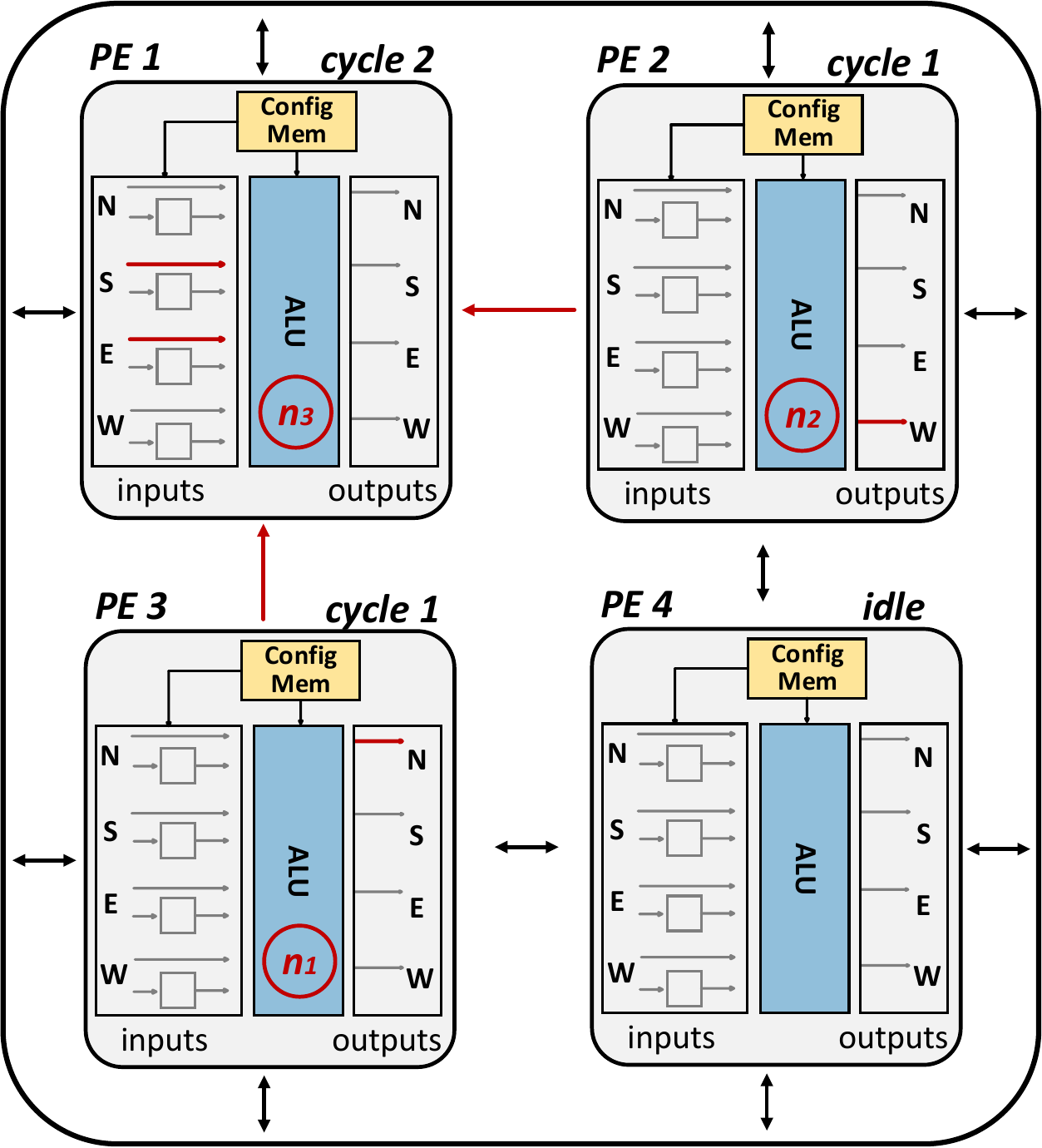}
    \caption{Detailed routing example of the highlighted sub-DFG in Figure~\ref{fig:motivation_dfg} on the 2$\times$2 CGRA from Figure~\ref{fig:motivation_cgra}. }
    \label{fig:motivation_routing}
\end{figure}

Figure~\ref{fig:motivation_routing} shows the communication, specifically the data routing between three PEs of a CGRA, for the red-highlighted portion of the DFG in Figure~\ref{fig:motivation_dfg}. In the first cycle, nodes n1 and n2 are executed on PE3 and PE2, respectively, as shown in the figure. The outputs generated are then routed towards PE1: PE3 sends its output to the north, and PE2 sends its output to the west. PE1 receives this data and processes it in the subsequent cycle to produce a new output, which is then sent to the next neighbouring PE.

This example illustrates two main problems of the traditional execution paradigm. First, only a small portion of the network is activated for communication for internal data dependencies of the highlighted sub-DFG. If dependent nodes are placed on neighbouring PEs, such simple and immediate routing does not need powerful routers. These routers benefit from distant routing when we cannot place dependent nodes within such ``proximity''. Nevertheless, a significant portion of the datapath is only fully engaged when there is extremely high routing congestion at the PE. Overall, in each cycle, only certain parts of the datapath are actively in use, while the other parts serve no functionality.  For the whole execution, this overprovisioned communication system does not provide a first-order benefit to the compute efficiency. 



Second, as the PE typically executes one node at a time, such individual execution comes with a high cost of programmability. For example, a simple connection in Figure~\ref{fig:motivation_routing} needs to configure two routers. This programmability incurs a significant encoding overhead and leads to unnecessarily complex solutions for simple problems. As shown in Figure~\ref{fig:intro_power}(a), communication configuration memory and the router consume 29\% and 15\% of power, respectively.
In summary, traditional CGRA designs often fail to align the provisioning of communication with the compute capabilities, and individuality in execution leads to a high cost of programmability, accentuating the inefficiency of resource misalignment. 



Through our work, we gain insights into the DFG behavior and identify patterns of communication, or motifs, between collective DFG nodes. We observe that the exhaustive communication possibilities between small groups of DFG nodes can be effectively implemented together in hardware using lightweight primitives. This approach enables hierarchical execution, with the architecture and mapper designed to exploit these motifs and their communication at two levels of granularity, and does not compromise the generality.

\section{Hierarchical Execution with Motifs }

This section first provides insight into harnessing simple connections in DFG with collective routing. Then, we discuss the design trade-off for collective routing and demonstrate the execution of the structural motifs with collective routing.

\subsection{Collective Routing}

DFGs are generated from the program by the compiler and represent the corresponding data dependencies; thereby, they do not have entirely arbitrary connections. For example, a single DFG node typically has only two inputs, indicating a limited degree of connection between multiple such nodes.
Supposing each node takes a maximum of two inputs, a DFG with $n$ nodes has a maximum of $2n$ edges. The limited number of inputs indicates the prevalence of relatively simple connections among DFG nodes, despite the fact that complex data dependencies also exist in the partial DFG.

These relatively simple connections raise an opportunity to handle the routing among multiple nodes together with a nimble design, instead of using multiple complex routers among multiple compute units. 
However, simple data dependencies in these sub-DFGs still have distinct connection patterns. To maintain versatility and support different patterns, we cannot resort to fixed connections in the hardware.  

\begin{figure}[!t]
  \centering
  \begin{minipage}{0.2\textwidth}
      \centering
    \includegraphics[width=\linewidth, page=7]{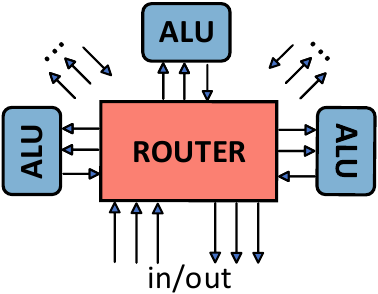}
    \caption{Prototype of collective routing}
    \label{fig:prototype}
  \end{minipage}
  \hfill 
  \begin{minipage}{0.25\textwidth}
     \centering
    \includegraphics[width=\linewidth,trim={0 -1cm 0 0cm}, page=4]{content/figures/Plaid.pdf}
    \caption{Fundamental motifs for three-node sub-DFGs}
    \label{fig:motif_example}
  \end{minipage}

\end{figure}

Figure~\ref{fig:prototype} shows a prototype of the proposed collective routing scheme: a single router provisions the input/output for a group of ALUs and connects with other such routers. Multiple ALUs execute nodes from a sub-DFG, and a router handles simple data dependencies within the sub-DFG. For example, for the sub-DFG in Figure~\ref{fig:motivation_dfg}, we can use three ALUs to execute and a router to handle related data dependencies. 

\textbf{\textit{Structural motifs}} refer to a set of sub-DFGs, which have the same number of DFG nodes and simple internal connections, such as the sub-DFGs shown in Figure~\ref{fig:motif_example}.  These motifs are naturally suitable for collective routing: instead of using multiple routers among multiple functional units, we can use a single router to route relevant data dependencies among multiple DFG nodes (functional units).  Hence, we can use minimal hardware to route the data dependencies of a motif and achieve very high utilization of the router. 

There are two challenges to leverage motifs for collective routing: The \textit{first challenge} involves determining the optimal number of ALUs (also the number of nodes in the motif)  connected to one router for designing the architecture. The hardware cost, including the router and its configuration, escalates significantly as the number of ALUs increases. Moreover, the complex connections among certain nodes in larger sub-DFGs make accommodating all connections with one router impractical.

The \textit{second challenge} is to design a compiler to effectively leverage the collective routing to execute the whole DFG. Although it is possible to disregard the fundamental hardware changes and continue mapping nodes individually, the true potential of these units is realized through more localized routing. In other words, we need an effective compiler to automatically identify motifs and judiciously schedule them to utilize the hardware unit of collective routing. Therefore, addressing this issue requires a comprehensive approach that considers both hardware and software perspectives to develop an efficient collective routing mechanism.



\subsection{Three-node Structural Motif}

The selection of an optimal motif size, which can benefit from local, collective routing, determines the number of ALUs that should be connected to a single router. A two-node sub-DFG has a fixed, single communication pattern between the nodes. Furthermore, as most DFG nodes have two inputs, the two-node sub-DFG has only one internal connection and needs more communication with the other routers, rendering it similar to individual execution and not feasible to be a motif.

Conversely, larger motifs do not appear frequently in DFGs and can lead to unnecessary fragmentation. Complex connections in certain DFG nodes complicate the identification of large sub-DFGs with straightforward data dependencies.
Moreover, these larger motifs tend to fragment the DFG into smaller pieces if not in the motif, diminishing the advantages of collective routing. As we mentioned earlier, accommodating larger communication patterns between nodes escalates hardware costs. Therefore, larger motifs are generally unsuitable for collective routing.

The three-node motif represents the smallest possible sub-DFG that benefits from reconfigurable communication patterns, i.e., using a router instead of a fixed connection. It aligns perfectly with the two inputs for most DFG nodes. As the smallest reconfigurable motif, it maximizes the number of motifs that can be partitioned from a given DFG and minimizes the number of standalone nodes. This balance makes the three-node motifs ideal for leveraging local, collective routing while maintaining hardware efficiency and reducing complexity. 
Prior work on patterns in much more complex and random real-world graphs and networks, have similar observations about the recurrence of three-node motifs~\cite{wernicke2006efficient, milo2002network}, affirming our approach.

Here, we formalize the solution to better understand the three-node motif. We start by examining Directed Acyclic Graphs (DAGs) with $N$ vertices.
We will build our understanding incrementally by considering the cases where the number of vertices, $|N|$, increases.

\noindent\textbf{Base case $|N| = 2$: }
For $|N| = 2$, the only possible DAG is $G_2: \{ (n_1, n_2) \} $. This represents a DFG where node $n1$ sends data to node $n2$, denoted as $n_1 \rightarrow n_2$.

\noindent\textbf{$|N| = 3$}: consider a new vertex $n_3$ added to $G_2$. The exhaustive set of possible DAGs, $G_{3i}$, that can be created by incorporating the new vertex $n_3$ can be enumerated as follows (visualized in Figure~\ref{fig:motif_example}):

\begin{itemize}[leftmargin=*]

    \item Node $n_3$ receives the input from $n_1$ , $ 
    E = \{(n_1, n_2), (n_1, n_3)\}$, called \textit{\textbf{fan-out} motif}.
    \item Node $n_3$ sends its output to $n_2$, $E = \{(n_1, n_2), (n_3, n_2)\}$, called \textit{\textbf{fan-in} motif}.
    \item Node $n_3$ sends its output to $n_1$, or $n_2$ sends its output to $n_3$: both are sequential chains,    $ E = \{(n_1, n_2), (n_2, n_3)\} $ or  $ E = \{(n_3, n_1), (n_1, n_2)\}$, called \textit{\textbf{unicast} motif}.
    \item Acyclic triangle: $ n_3 \rightarrow n_1 \rightarrow n_2 \leftarrow n_3$, 
   $ E = \{(n_3, n_1), (n_3, \\n_2),  (n_1, n_2)\} $.  The acyclic triangle can be derived from any of these three basic motifs by introducing a single additional edge, and thus is not a basic motif. 
\end{itemize}
\noindent The first three basic motifs here serve as exhaustive, fundamental building blocks from which any other DFG with $|N| = 3$ can be constructed.

\begin{figure}[!t]
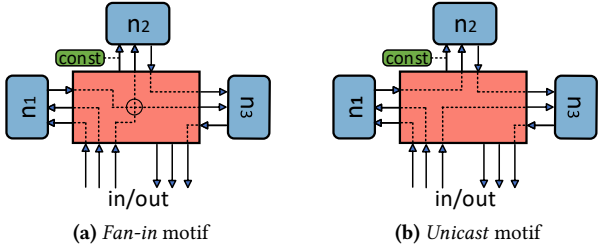
%
    \centering
    \subfloat[\textit{Fan-in} motif ]{{\includegraphics[width=.42\linewidth, trim={0 0 0 0.5cm},page=8]{content/figures/Plaid.pdf} }}%
    \qquad
    \subfloat[\textit{Unicast} motif]{{\includegraphics[width=.42\linewidth, trim={0 0 0 0.5cm},page=9]{content/figures/Plaid.pdf} }}%
    \caption{Collective routing for \textit{fan-in} and \textit{unicast} motifs.}%
    \label{fig:motif_execution}%
\end{figure}

\begin{figure*}[!t]
  \centering
  
    \includegraphics[width=0.9\linewidth, page=11]{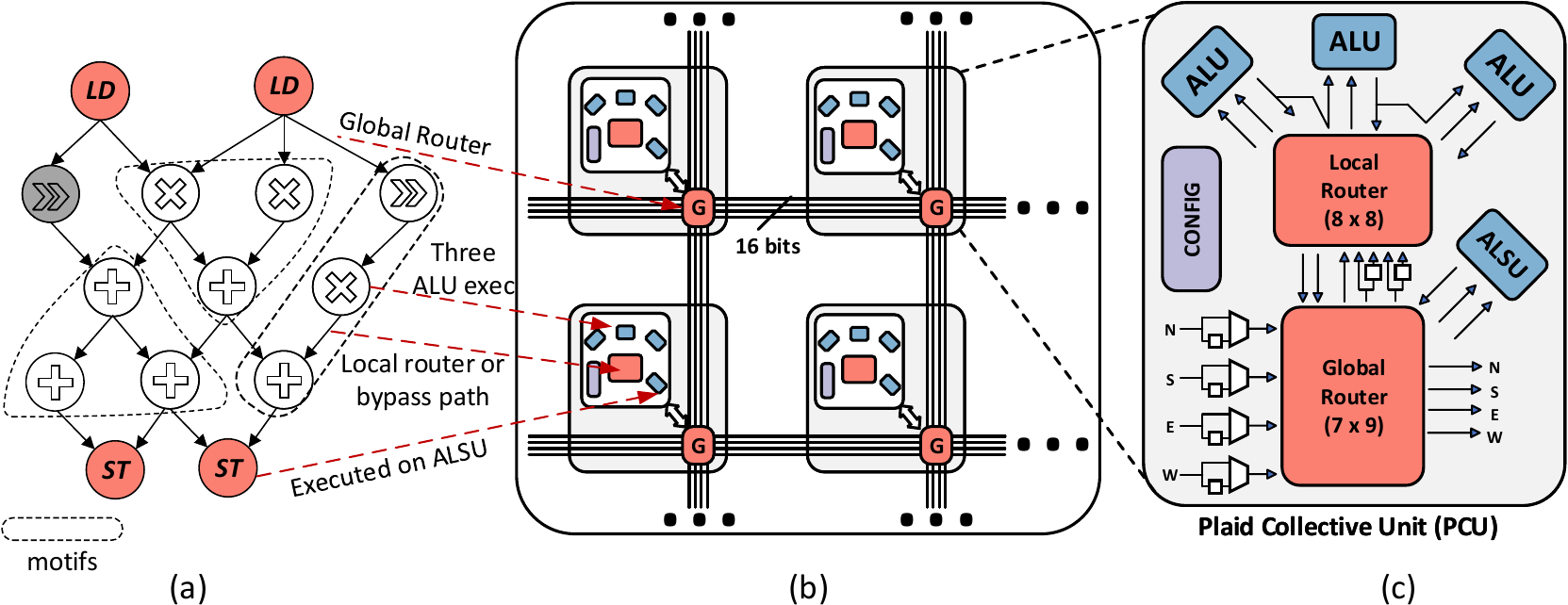}
\caption{ (a) Hierarchical DFG with multiple motifs. (b) Plaid architecture overview (c) Plaid Collective Unit}
\label{fig:motif_microarchitecture}
\end{figure*}

\noindent\textbf{Hierarchical composition:}
For a DAG, given all possible subgraphs for \(|N| = 3\), any graph with \(3 \times n\) nodes can be constructed using the above subgraphs as interconnected building blocks. This approach leverages the principle of graph composition, where larger graphs are formed by combining smaller, well-understood subgraphs~\cite{graphComp1, graphComp2}. We can easily prove that, for DFGs \(G_{3n+k}\), with \(3n + k\) nodes, where \(0 < k < 3\), the graph can be constructed by using \(\left\lfloor \frac{3n + k}{3} \right\rfloor = n\) subgraphs of 3 nodes (three basic motifs), supplemented with \(k\) additional standalone helper nodes. Formally,  $ G_{3n+k} = \left(\bigcup_{i=1}^{n} G_{3i}\right) \cup H_k \cup E'' $, where \(H_k\) represents the standalone nodes, and \(E''\) represents the additional edges that connect different subgraphs and various standalone nodes with each other. A DFG can thus be constructed by utilizing the three motifs and individual standalone nodes wherever required.

Figure~\ref{fig:motif_execution} demonstrates the execution of \textit{fan-in} (which forms the highlighted section in Figure~\ref{fig:motivation_dfg}) and \textit{unicast} motifs using three ALUs and a single router. The internal data dependencies here are managed within the router. For instance, communications such as \(n_1 \rightarrow n_2\) and \(n_2 \rightarrow n_3\) within the unicast motif are routed locally. Other non-constant inputs and outputs, which form the standalone nodes \(H_k\) or additional edges \(E''\), engage the local router to communicate with other routers. This arrangement ensures high router utilization and alleviates the demand for global communication by routing internal dependencies locally.

For a better understanding of the hierarchical nature, Figure~\ref{fig:motif_microarchitecture}(a) presents a DFG example featuring multiple motifs.  We assume the memory access unit is not connected to the collective router and explain the execution of memory operations with a hierarchical on-chip network later. It happens that some standalone nodes are not included in the motif. Such a node can still be executed on any ALU, without the loss of generality. The collective routing mechanism still applies for such nodes, but the corresponding data dependencies might not be routed locally. 
These motifs enable a \textit{\textbf{hierarchical execution}} on the CGRA. To accommodate the high efficiency of the hierarchical execution, we first need to manage the communications among these motifs. As we collectively route these relatively simple dependencies, the communications among motifs are substantial.  
Second, we need to reorganize the entire CGRA to enhance overall execution efficiency. 

\section{Plaid Architecture}


The \textit{Plaid} CGRA is designed with two levels of granularity in execution as illustrated in Figure~\ref{fig:motif_microarchitecture}(b). Each tile executing a motif is called the Plaid Collective Unit (PCU). A PCU incorporates the collective routing circuitry and executes individual motifs, i.e., handles simple DFG connections. Multiple PCUs are interconnected in a mesh network and interact with each other to facilitate overall DFG execution, supporting the communication between various motifs and standalone nodes, including the complex connections in the DFG. Hierarchical execution enables the exploiting of patterns in communication and achieves high efficiency for \textit{Plaid}. 



\subsection{Plaid PCU}

The \textit{Plaid} PCU is specifically designed to execute three-node motifs. The collective routing prototype, as discussed previously, has been adapted for this PCU’s micro-architecture, featuring three ALUs (motif compute unit) connected to a local router. Figure~\ref{fig:motif_microarchitecture}(c) depicts the internal connections within the PCU. The ALUs, which are 16-bit compute units, support ADD, MUL, SHIFT, and various bit-wise operations, totalling 15 operations. 
The local router delivers inputs to each of the three ALUs per cycle and collects outputs from them, enabling the execution of all possible three-node motifs.

Although the collective routing prototype can handle all three node motifs with the identically positioned ALUs around the router, we add \textit{virtual} bypass paths between the ALUs. 
These bypass paths provide shorter routing paths and are prioritized by the mapper, which typically favours an execution sequence from the left-most to the right-most ALU. However, this execution order is not obligatory. We identify that adhering strictly to this order for all motifs can lead to resource under-utilization. To address this, we developed more flexible scheduling options, which we will detail in Section~\ref{sec:mapping}. Nevertheless, these bypass paths are still beneficial as they reduce the routing pressure on the local router. 

\subsection{Hierarchical Network-on-Chip} 
The local router connects to a global router, which facilitates communication between motifs across different PCUs and links motifs to the data memory. The global router is connected to an Arithmetic-Load-Store Unit (ALSU), which is integral to the PCUs along the edges interfacing with the data memory. The ALSU manages load-store operations within the DFG via a dedicated datapath to the data memory. Additionally, it facilitates the mapping of predication operations and standalone nodes facing routing challenges.

The global router has ports facing the N, S, E, and W directions, enabling inter-motif (inter-PCU) communication.
The architecture creates two distinct datapaths: a global datapath, which connects all PCUs across the CGRA through the global router, functioning like a data conveyor belt; a local datapath within each PCU, managed by the local router, which aims to handle all communication internally. Yet, when necessary, it interacts with the global path, either retrieving or depositing data onto the conveyor belt. This hierarchical organization ensures efficient and flexible data management, optimizing both local and global communication within the system.


\noindent\textbf{Considerations for the Global Datapath}: The paths between the global and local routers include options for temporal data buffering using registers, as shown in Figure~\ref{fig:motif_microarchitecture}(c). 
Data that continuously cycles from the global router to the local and back again can potentially get stuck in a loop when synthesized. To prevent the formation of such hardware loops, we enforce constraints within the interconnects in the compiler and the hardware. The compiler constraints are designed to block any configuration that could create a closed loop. The hardware constraints ensure no such datapath can be created. These constraints are verified post-synthesis through the EDA tool, which can report any such violations.

\subsection{Configuration} With each PCU now managing three ALUs, one ALSU, two routers, and multiple registers along the datapath, the configuration memory encoding bits need a comprehensive redesign. The routers alone consume about half of these encoding bits, highlighting their role in the system. The ALUs connected to the local router require eight-bit constants and four-bit fields for operation selection per cycle. 
Each instruction, or configuration entry, comprises a total of 120 bits, including the local and global configuration space.

\subsection{Efficient Domain-Specialization}
\label{sec:domain_spec}

Recent advancements in CGRAs have enabled domain-specific optimizations~\cite{revamp, melchert2023apex, tan2023vecpac} in the architecture. 
Plaid's execution paradigm, which already exploits communication patterns, can further enhance efficiency through domain-specific specialization. We exploit the existence of recurrent motifs in a domain and specialize a few PEs in architecture that are domain-optimized for those motifs. Essentially, we hardwire a fixed motif inside certain PEs to replace the local router while maintaining the full reconfiguration capability of the global datapath. This approach reduces the configuration space for these specific PEs, compelling the mapper to prioritize them for mapping the corresponding motifs they were specialized for.

\section{Plaid Compiler}
\label{sec:mapping}
This section first formulates the mapping problem and then introduces our algorithm for generating motifs, followed by the introduction of the mapping algorithm.

\subsection{Mapping Problem Formulation}
Given a DFG and a CGRA, application mapping is executed through modulo scheduling. This technique allows for a new loop iteration at every initiation interval (II). Initially, we ascertain the lower bound of II, designated as the Minimum II (MII). This is determined by taking the greater of two metrics: the resource minimum II (ResMII) and the recurrence minimum II (RecMII) caused by inter-iteration data dependency. For each prospective II value, a time-extended resource graph of the CGRA, corresponding to II cycles, is constructed. This graph is referred to as the Modulo Routing Resource Graph (MRRG). Owing to the cyclical nature of the schedule, which repeats every II cycle, there is connectivity between the resources at cycle II-1 and those at cycle 0 within the MRRG—hence the term “modulo.” The compiler aims to achieve an optimal mapping of the DFG onto the MRRG, aiming to minimize the II value.

Given a DFG $D = (V_D, E_D)$ and a \textit{Plaid} CGRA instance, the problem is to generate a hierarchical DFG with multiple motifs and standalone nodes, $HD = ( M_{HD}, E_{HD})$ (standalone node is a special motif where motif node number is one), and construct a minimally time-extended MRRG of the \textit{Plaid} instance $P_{II} =
(V_{P}, E_P)$ , which has a valid mapping $\phi = (\phi M , \phi E)$ from $HD$.

\subsection{Mapping DFG onto CGRA}

\begin{algorithm}[!t]
\small
\caption{Motif Generation}
\label{alg:motif}

\KwIn{DFG;}
\KwOut{Hierarchical DFG with motifs}
\SetKwFunction{detectMotif}{detectMotif}

Generate the initial motifs greedily\;

\While{the motif number increases}{
    Randomly break down one motif\;
    Randomly sort standalone nodes\;
    \ForEach{standalone node}{
         \uIf{find a motif pattern with this node}{
            Generate the motif and update standalone nodes\;
          }
    }
}

\setlength{\abovedisplayskip}{-15pt}
\setlength{\textfloatsep}{0pt}

\end{algorithm}

\begin{figure}[t]
   \centering
    \includegraphics[width=.9\linewidth, page=17]{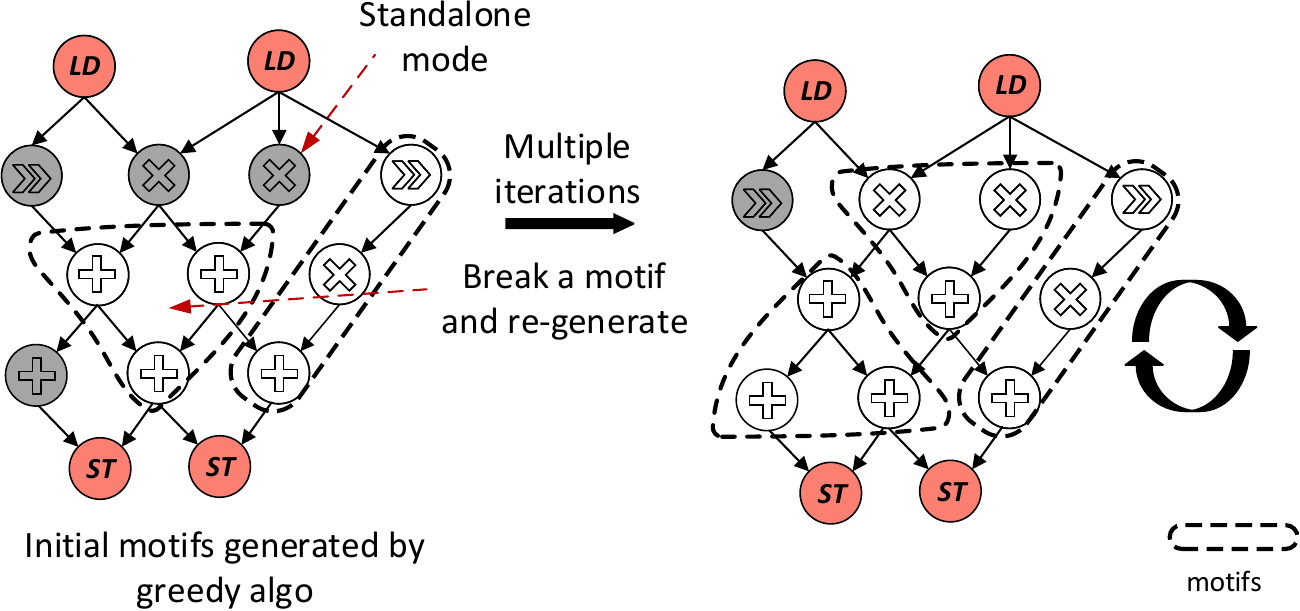}
    \caption{Example with Algorithm~\ref{alg:motif}}
    \label{fig:compiler_motif_gen_example}
\end{figure}

Algorithm~\ref{alg:motif} delineates a systematic approach to identify motifs ($M_{HD}$) to generate a hierarchical DFG ($HD$). 
This algorithm is easily extendable to the motif with any number of nodes.
Figure~\ref{fig:compiler_motif_gen_example} shows an example using Algorithm~\ref{alg:motif}. 
This process starts with the greedy generation of initial motifs, leveraging the fundamental set of motif structure patterns to traverse the DFG and generate motifs (line 1). However,  as shown in Figure~\ref{fig:compiler_motif_gen_example}, the greedy generation is sub-optimal and there are many standalone nodes in the initial generation.

We adopt an iterative algorithm to improve motif identification. The algorithm randomly deconstructs a motif, randomly sorts standalone nodes, and traverses the DFG starting from standalone nodes to find new motifs as long as they match the motif pattern (lines 3-7). For example, in Figure~\ref{fig:compiler_motif_gen_example}, we break the left motif, randomly sort standalone nodes, and try to find sub-DFGs that match motif patterns to generate more motifs. 
We keep running the deconstruction and re-generation process until the number of motifs does not increase (lines 2-7), or the number of motifs exceeds standalone nodes. The latter is to ensure the utilization of the motif compute unit and ALSU in PCU. 
The method’s iterative nature helps refine motif integration within the DFG, aiming for optimized motif generation.

\begin{algorithm}[!t]
\small
\caption{Hierarchical mapping }
\label{alg:mapping}

\KwIn{Hierarchical DFG, CGRA architecture description;}
\KwOut{Mapping}

Sort motifs by data dependency\;

\While{Mapping is not valid}{
    \ForEach{motif} {
        Map the motif to a PE with the least routing resource\;
    }
    
    \While{not find a valid mapping or exceed time limitation}{
        Unmap one motif \;
        
           Randomly select one placement candidate\;
           Generate different motif schedules\;
            \ForEach{motif schedule}{
               Route this motif’s operands and dependencies to
        the network using Djkstra’s algorithm\;
            
            Select the combination yielding the highest objective\;
        }
        
    }
    II =  II + 1 \;
}
\setlength{\abovedisplayskip}{-15pt}
\setlength{\textfloatsep}{0pt}

\end{algorithm}

Algorithm~\ref{alg:mapping} presents the hierarchical mapping to efficiently map DFGs with identified motifs, generated by the Algorithm~\ref{alg:motif},  onto a Plaid CGRA. 
Algorithm~\ref{alg:mapping} augments simulated annealing~\cite{wijerathne2022morpher,chin2017cgra} to hierarchically schedule motifs. Figure~\ref{fig:motif_microarchitecture} shows the corresponding hardware unit to handle motifs, internal data dependencies, and global communication.
We use a cost function to guide the mapping process, which includes metrics such as the number of unmapped nodes ($V_D$), congestion levels, and the usage of routing resources.

\begin{figure}[t]
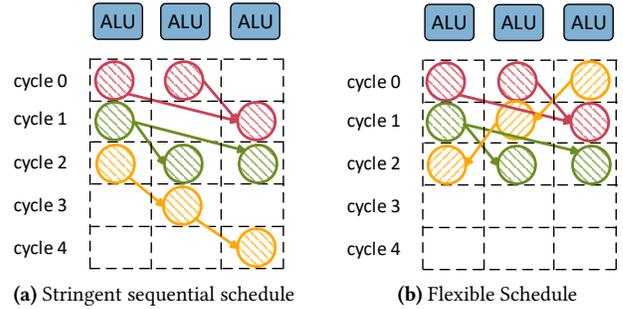
%
    \centering
    \subfloat[Stringent sequential schedule]{{\includegraphics[width=.43\linewidth, trim={0 0 0 0cm},page=12]{content/figures/Plaid.pdf} }}%
    \qquad
    \subfloat[Flexible Schedule]{{\includegraphics[width=.43\linewidth, trim={0 0 0 0cm},page=13]{content/figures/Plaid.pdf} }}%
    \caption{Schedules for three motifs. The right one has a higher resource utilization.}%
    \label{fig:motif_compiler_schedule}%
\end{figure}

Initially, motifs within the DFG are sorted according to the data dependency (line 1), which prioritizes critical paths for early mapping.
The algorithm starts with the MII.  If no valid mapping is found at the current II, it increments the II by one, continuing this process until it either finds a valid mapping or exceeds the maximum II determined by the configuration memory size (lines 2-12).
In each II, initially, sorted motifs are mapped onto the hardware unit $V_{P}$ with the least routing resources (lines 3 and 4). If a valid mapping is not achieved within the initial step, the algorithm uses an iterative methodology to generate a valid mapping (lines 4-10). In each iteration, one motif ($M_{HD}$) is unmapped (line 6), and we randomly select one motif unit (line 7) as the placement candidate. We can place a single node (special $M_{HD}$) onto any functional unit $V_{P}$ if it supports the operation.

The motif compute unit's bypass connection does not restrict the node placement within a motif from left to right. For example, in the unicast motif, the first node and the second node  are not required to be placed onto the leftmost ALU and the middle ALU, respectively. Instead, the first node can be placed onto the rightmost ALU. Of course, the bypass connection is not utilized in such case.  A stringent placement can cause motif compute units to be under-utilized if any ALU is not available.

Figure~\ref{fig:motif_compiler_schedule} illustrates two distinct schedules for three motifs on the motif compute unit. In Figure~\ref{fig:motif_compiler_schedule}(a), the scheduling order is strict to the above-mentioned sequential order: the first node is allocated to the leftmost ALU, the second node to the middle ALU, and the third node to the rightmost ALU. However, this stringent scheduling sequence leads to an under-utilization of resources. In contrast, Figure~\ref{fig:motif_compiler_schedule}(b) depicts a more flexible scheduling approach, where the \textit{unicast} motif adopts a `reversed' sequence compared to that in Figure~\ref{fig:motif_compiler_schedule}(a), thereby enhancing the utilization of the collective unit.

To support flexible scheduling, we generate several schedule templates for each motif type. For example, for the fan-out motif, we have the following schedule templates: \\$\{(n1,c), (n2,c+1), (n3,c+1)\}$,  $\{(n1,c), (n2,c+1), (n3,c+2)\}$,  $\{(n1,c), (n2,c+2), (n3,c+1)\}$, $\{(n3,c+1), (n2,c+1), (n1,c)\}$, $\{(n3,c+2), (n2,c+1), (n1,c)\}$, \& $\{(n3,c+1), (n2,c+2 ), (n1,c)\}$. Each template comprises three tuples representing an ALU in left-to-right order, with each tuple containing the node in motif and its scheduled cycle. These templates ensure sufficient flexibility and obviate exhaustive searches on the motif compute unit. 
For the placement candidate, we use Dijkstra’s algorithm to assess each schedule (line 10). The schedule that yields the highest objective value (the least cost) is selected (line 11). Like typical simulated annealing, we can occasionally accept a "worse" movement to overcome the local minimum. We repeat this process until we find a valid mapping or exceed the mapping time limitation.

\section{Experimental Methodology}
We describe the setup and methodology to evaluate the end-to-end framework for \textit{Plaid} in this section.

\subsection{Architecture Synthesis}
We implement the \textit{Plaid} architecture using Verilog RTL and synthesize it with Cadence Genus targeting a 22nm FDSOI technology node at 100MHz. The power consumption metrics are obtained from post-synthesis estimates provided by the synthesis tool.
\textit{Plaid} features a 2$\times$2 PCU array and four 4KB data memory banks, with each PCU having a 16-entry configuration memory. As each PCU has four functional units, the 2$\times$2 \textit{Plaid} has the same number of functional units as typical 4$\times$4 CGRA. Moreover, we also evaluate the scaled 3$\times$3 \textit{Plaid} with the same number of functional units as a 6$\times$6 CGRA.

\subsection{Compiler and Execution}
The \textit{Plaid} framework processes annotated C code, as shown in Figure~\ref{fig:motivation_dfg}, and generates static configurations for the \textit{Plaid} CGRA using the Morpher toolchain\cite{wijerathne2022morpher}, which includes a mapper and a cycle-accurate C++ simulator. 
The compiler typically maps the kernel in a few minutes. 
As typical CGRAs are statically scheduled, the performance is deterministic and known at compilation time. With the II and the total number of loop iterations, we can precisely calculate the overall performance (cycles). The primary purpose of the simulation is to verify the mapping and hardware design.


The host processer loads the CGRA configuration bits generated by the compiler to the CGRA fabric and sends input data to the SPM. Then,  it triggers the CGRA to fetch the configuration, load the data, perform computations, and store the results back in the SPM. Spatio-temporal CGRA can read the configuration memory per cycle in a modulo way. 
Ultimately, data from the SPM is moved back to the main memory for further processing by the host processor.

\begin{table}[!t]
\centering
\footnotesize
\renewcommand{\arraystretch}{0.9}
\begin{tabular}{|>{\centering\arraybackslash}m{0.95cm}|>{\centering\arraybackslash}m{0.8cm}|>{\centering\arraybackslash}m{0.8cm}|>{\centering\arraybackslash}m{0.8cm}|>{\centering\arraybackslash}m{0.8cm}|>{\centering\arraybackslash}m{0.8cm}|>{\centering\arraybackslash}m{0.8cm}|}
\hline
\rowcolor[HTML]{C0C0C0} 
\multicolumn{1}{|c|}{\cellcolor[HTML]{C0C0C0}\textbf{domain}} & \multicolumn{1}{c|}{\cellcolor[HTML]{C0C0C0}\textbf{kernel}} & \multicolumn{1}{c|}{\cellcolor[HTML]{C0C0C0}\textbf{unroll}} & \multicolumn{1}{c|}{\cellcolor[HTML]{C0C0C0}\textbf{char$^1$}} & \multicolumn{1}{c|}{\cellcolor[HTML]{C0C0C0}\textbf{kernel}} & \multicolumn{1}{c|}{\cellcolor[HTML]{C0C0C0}\textbf{unroll}} & \multicolumn{1}{c|}{\cellcolor[HTML]{C0C0C0}\textbf{char}} \\ \hline
\multicolumn{1}{|c|}{} & atax & 2 & 15,6,6 & atax & 4 & 27,14,11 \\ \cline{2-7} 
\multicolumn{1}{|c|}{} & bicg & 2 & 23,11,10 & bicg & 4 & 42,23,19 \\ \cline{2-7} 
\multicolumn{1}{|c|}{} & doitgen & 2 & 18,9,9 & doitgen & 4 & 34,21,10 \\ \cline{2-7} 
\multicolumn{1}{|c|}{} & gemm & 2 & 21,12,12 & gemm & 4 & 37,24,23 \\ \cline{2-7} 
\multicolumn{1}{|c|}{} & gemver & 2 & 21,11,10 & gemver & 4 & 41,23,19 \\ \cline{2-7} 
\multicolumn{1}{|c|}{\multirow{-6}{*}{\begin{tabular}[c]{@{}c@{}}Linear\\  Algebra\end{tabular}}} & gesumm & 2 & 22,9,8 & gesumm & 4 & 38,19,16 \\ \hline
 & conv2x2 & 1 & 20,12,10 & conv3x3 & 1 & 37,26,17 \\ \cline{2-7} 
 & dwconv & 1 & 7,3,2 & dwconv & 5 & 31,19,13 \\ \cline{2-7} 
\multirow{-3}{*}{\begin{tabular}[c]{@{}l@{}}Machine\\  Learning\end{tabular}} & fc & 1 & 17,8,7 &  & \multicolumn{1}{l|}{} &  \\ \hline
 & cholesky & 2 & 14,5,4 & cholesky & 4 & 28,11,8 \\ \cline{2-7} 
 & durbin & 2 & 14,7,4 & durbin & 4 & 28,15,8 \\ \cline{2-7} 
 & fdtd & 2 & 16,7,6 & fdtd & 4 & 32,15,12 \\ \cline{2-7} 
 & gramsc & 2 & 15,5,4 & gramsc & 4 & 25,11,8 \\ \cline{2-7} 
 & jacobi & 1 & 16,7,5 & jacobi & 2 & 30,15,12 \\ \cline{2-7} 
 & jacobi & 4 & 54,30,27 & seidel & 1 & 22,11,9 \\ \cline{2-7} 
\multirow{-7}{*}{Image} & seidel & 2 & 44,23,21 &  &  &  \\ \hline
\end{tabular}
\begin{tablenotes}
    \item[a]\scriptsize{1. Characteristics contain the number of DFG nodes, the number of compute nodes, and the number of compute nodes covered by motifs.}
\end{tablenotes}
\caption{Evaluated workloads}
\label{tab:meth_kernels}
\end{table}

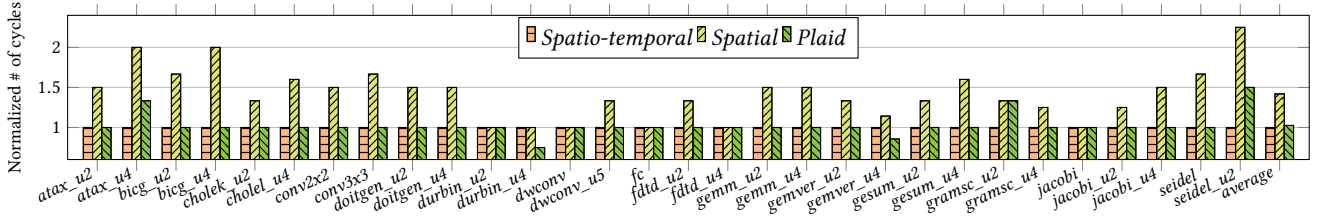
\begin{figure*}[t]
	\scriptsize
	\centering
	\begin{tikzpicture}

\pgfplotsset{set layers}
\begin{axis}[
width=21cm, 
height=3.5cm,
ylabel style={align=center},
ylabel={Normalized \#  of cycles},
bar width=3.5pt,
ybar=0pt,
xtick=data,
xticklabels from table={content/data/perf.txt}{Benchmark},
x tick label style={rotate=25,anchor=east},
ymajorgrids=true,
yminorgrids=true,
yticklabel={$\pgfmathprintnumber{\tick}$},
legend columns=3,
legend style={at={(0.5,0.7)}, anchor=south, font=\small},
enlarge x limits = 0.025,
 x post scale = 0.85
]

\addplot[fill=Apricot  ,  postaction={pattern=horizontal lines}]
table[
x expr = \coordindex,
y = hycube,
]{content/data/perf.txt};

\addlegendentry{\em Spatio-temporal}

\addplot[fill=GreenYellow,postaction={pattern=north east lines}]
table[
x expr = \coordindex,
y = Spatial	,
]{content/data/perf.txt};

\addlegendentry{\em Spatial}

\addplot[fill=LimeGreen  ,  postaction={pattern=north west lines}]
table[
x expr = \coordindex,
y = plaid	,
]{content/data/perf.txt};
\addlegendentry{\em Plaid}

\end{axis}

\end{tikzpicture}
	\caption{Performance of \textit{Plaid} and a spatial CGRA normalized to a spatio-temporal CGRA. \textit{u2} :unrolling factor of 2. }
	\label{fig:eval_perf_comp}
\end{figure*}

\subsection{Baseline CGRAs}
Given the significant redesign of the CGRA architecture, we benchmark our results against several baselines:\newline
\noindent\textbf{Spatio-temporal CGRA}, like typical CGRAs \cite{mei2003adres, wang2019hycube, dave2018ramp, tan2023vecpac} as shown in Figure~\ref{fig:motivation_cgra},  has a 4$\times$4 PE array with a mesh network and the same SPM configuration with  2$\times$2 \textit{Plaid}. Each PE has 16-entry configuration memory.
\newline
\noindent\textbf{Spatial CGRA} follows the state-of-the-art energy-efficient spatial CGRAs \cite{gobieski2021snafu, gobieski2022riptide, serafin2023pipestitch} but with a mesh network. The spatial CGRA  has a 4$\times$4 PE array. Mapping complex kernels ($II>1$) onto spatial CGRAs requires partitioning the DFG into several DFGs. We develop a Python script to partition DFGs. Additional loads and stores are introduced during partition to put intermediate data in SPM.  
\newline
\noindent\textbf{Domain-optimized spatio-temporal CGRA} is based on the spatio-temporal CGRA and optimized for the machine learning domain~\cite{revamp}. This lowers the performance of the spatio-temporal CGRA for other domains but improves the energy and area efficiency of the target domain. 
\newline
\noindent Baselines are implemented using the same technology node for consistent area and power comparisons. We use two mappers for these baselines and select the one with higher performance. The first is PathFinder, adapted from~\cite{mcmurchie1995pathfinder,wijerathne2022morpher,tan2020opencgra}, consisting of around 3K lines of C++ code. The second mapper utilizes simulated annealing, as detailed in~\cite{chin2017cgra,wijerathne2022morpher,weng2020dsagen}, implemented with around 2K lines of C++ code.

\subsection{Workloads}

To demonstrate that \textit{Plaid} maintains the versatility of CGRAs, we benchmark our work using a diverse set of kernels from various applications and benchmark suites, as detailed in Table~\ref{tab:meth_kernels}.  We unroll DFGs to evaluate the capability of \textit{Plaid} to handle complex data dependencies, leading to 30 DFGs in total.  We use five kernels from TinyML\cite{banbury2021mlperf} to represent typical machine learning workloads. 
For \textit{dwconv}, we unroll with 5 as the trip count is not divisible by other less unrolling factors. We evaluate the linear algebra benchmark suite and image process kernels from PolyBench~\cite{karimov2018polybench}, assessing the performance of kernels from this suite. To ensure a similar number of kernels from each domain, we use the first six kernels from the PolyBench linear algebra suite. For each DFG, Table~\ref{tab:meth_kernels} shows the number of nodes, compute nodes, and nodes covered by motifs. As motif compute unit does not provide memory access, all the nodes in motifs are compute nodes. Moreover, we also execute the two-node motif with the motif compute unit. As we provide flexible schedules, this does not affect the schedule of three-node motifs. 
\newline

\noindent\textbf{Application-level mapping:} to evaluate system-level performance, we evaluate three DNN applications adapted from TinyML~\cite{banbury2021mlperf}. These three DNN applications comprise 10, 13, and 16 layers, respectively. Most layers are Convolution layers and DepthWiseConv layers.  

\begin{figure}[t]
	\scriptsize
	\centering
	\pgfplotsset{
  compat=1.9,
}
\begin{tikzpicture}[align=center]
    \begin{axis}[
    xbar stacked,
    xmajorgrids = true,
    xmin=0,xmax=100,
    ytick style={draw=none},
    ytick = data, yticklabels = {Area},
    tick align = outside, xtick pos = left,
    bar width=7mm, y=13mm,
    every node near coord/.append style={font=\footnotesize},
    nodes near coords={\pgfmathprintnumber{\pgfplotspointmeta}\%}, 
    xticklabel={\pgfmathprintnumber{\tick}\%},
    nodes near coords align={center}, 
    enlarge y limits=0.6, 
    legend style={/tikz/every even column/.append style={column sep=0.18cm}},
    legend image post style={xscale=0.6, yscale=0.8},
    legend columns = 3,
    legend style={at={(0.5,1.15)},anchor=south},
    ]
        \addplot [fill=pie16] coordinates{(5,1)};
        \addplot [fill=pie15] coordinates{(11,1)};
        \addplot [fill=pie14] coordinates{(21,1)};
        \addplot [fill=pie13] coordinates{(24,1)};
        \addplot [fill=pie12] coordinates{(30,1)};
        \addplot [fill=pie11] coordinates{(9,1)};
        \legend{Local Router, Global Router, Config Compute, Config Communication, Compute, Others}
    \end{axis}
\end{tikzpicture}
	\caption{Area breakdown for Plaid's CGRA fabric}
	\label{fig:plaid_area}
\end{figure}
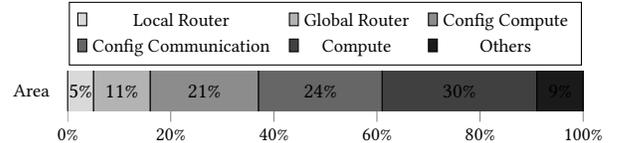

\section{Evaluation}




To highlight the overprovisioned communication in CGRAs, we compare CGRAs with equivalent theoretical throughput (the same number of functional units) from three perspectives: performance, energy consumption, and performance per area. Second, we showcase \textit{Plaid}'s scalability by evaluating larger architecture versions using a 3$\times$3 PCU array, to highlight \textit{Plaid} can maintain efficiency as it scales.  Third, to evaluate the effectiveness of our mapper, we compare it with popular CGRA mappers on \textit{Plaid}.  Finally, we implement a domain-optimized \textit{Plaid} and compare it with a domain-optimized (specifically machine learning) spatio-temporal CGRA.

\begin{figure*}[t]
	\scriptsize
	\centering
	\begin{tikzpicture}

\pgfplotsset{set layers}
\begin{axis}[
width=21cm, 
height=3.5cm,
ylabel style={align=center},
ylabel={Normalized \\ total energy},
bar width=3pt,
ybar=0pt,
xtick=data,
xticklabels from table={content/data/energy.txt}{Benchmark},
x tick label style={rotate=25,anchor=east},
ymajorgrids=true,
yminorgrids=true,
yticklabel={$\pgfmathprintnumber{\tick}$},
legend columns=3,
legend style={at={(0.5,0.75)}, anchor=south, font=\small},
enlarge x limits = 0.025,
 x post scale = 0.85
]

\addplot[fill=Apricot  ,  postaction={pattern=horizontal lines}]
table[
x expr = \coordindex,
y = hycube,
]{content/data/energy.txt};

\addlegendentry{\em Spatio-temporal}

\addplot[fill=GreenYellow,postaction={pattern=north east lines}]
table[
x expr = \coordindex,
y = Spatial	,
]{content/data/energy.txt};

\addlegendentry{\em Spatial}

\addplot[fill=LimeGreen  ,  postaction={pattern=north west lines}]
table[
x expr = \coordindex,
y = plaid	,
]{content/data/energy.txt};
\addlegendentry{\em Plaid}

\end{axis}

\end{tikzpicture}
	\caption{ Energy consumption of \textit{Plaid} and a spatial CGRA normalized to a spatio-temporal CGRA.}
	\label{fig:eval_energy_comp}
\end{figure*}
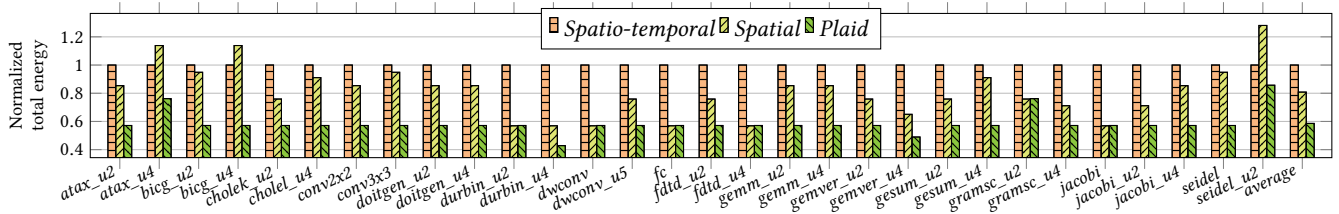

\begin{figure*}[t]
	\scriptsize
	\centering
	\begin{tikzpicture}

\pgfplotsset{set layers}
\begin{axis}[
width=21cm, 
height=3.2cm,
ylabel style={align=center},
ylabel={Normalized \\ Perf/Area},
bar width=3pt,
ybar=0pt,
xtick=data,
xticklabels from table={content/data/perf_per_area.txt}{Benchmark},
x tick label style={rotate=25,anchor=east},
ymajorgrids=true,
yminorgrids=true,
yticklabel={$\pgfmathprintnumber{\tick}$},
legend columns=3,
legend style={at={(0.54,0.8)}, anchor=south, font=\small},
enlarge x limits = 0.025,
 x post scale = 0.85
]

\addplot[fill=Apricot  ,  postaction={pattern=horizontal lines}]
table[
x expr = \coordindex,
y = hycube,
]{content/data/perf_per_area.txt};

\addlegendentry{\em Spatio-temporal}

\addplot[fill=GreenYellow,postaction={pattern=north east lines}]
table[
x expr = \coordindex,
y = Spatial	,
]{content/data/perf_per_area.txt};

\addlegendentry{\em Spatial}

\addplot[fill=LimeGreen  ,  postaction={pattern=north west lines}]
table[
x expr = \coordindex,
y = plaid	,
]{content/data/perf_per_area.txt};
\addlegendentry{\em Plaid}

\end{axis}

\end{tikzpicture}
	\caption{Performance per area of \textit{Plaid} and a spatial CGRA normalized to a spatio-temporal CGRA}
	\label{fig:eval_perf_per_area}
\end{figure*}
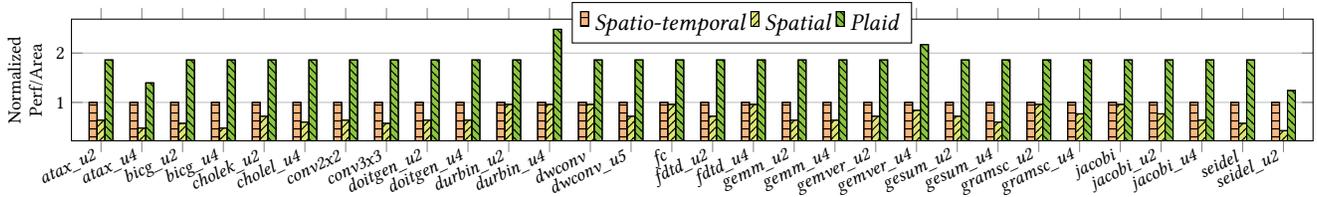

\begin{figure}
    \scriptsize
   \centering
    \subfloat [  Energy consumption]{
\begin{tikzpicture}
\pgfplotsset{set layers}
\begin{axis}[
width=4.8cm, 
height=3cm,
ylabel style={align=center},
ylabel={Normalized \\total energy},
bar width=4.5pt,
ybar=0pt,
xtick=data,
xticklabels from table={content/data/dnn_energy.txt}{Benchmark},
ymajorgrids=true,
yminorgrids=true,
yticklabel={$\pgfmathprintnumber{\tick}$},
legend columns=2,
legend style={at={(0.8,0.9)}, anchor=south, font=\small},
enlarge x limits = 0.2,
 x post scale = 0.75
]

\addplot[fill=LimeGreen]
table[
x expr = \coordindex,
y = efficiency	,
]{content/data/dnn_energy.txt};

\end{axis}
\end{tikzpicture}
\label{fig:performance_3x3}
 
}
\hfil
\subfloat  [Performance per area]{
\begin{tikzpicture}

\pgfplotsset{set layers}
\begin{axis}[
width=4.8cm, 
height=3cm,
ylabel style={align=center},
ylabel={Normalized  \\ Perf/Area},
bar width=4.5pt,
ybar=0pt,
xtick=data,
xticklabels from table={content/data/dnn_area.txt}{Benchmark},
ymajorgrids=true,
yminorgrids=true,
yticklabel={$\pgfmathprintnumber{\tick}$},
legend columns=2,
legend style={at={(0.8,0.9)}, anchor=south, font=\small},
enlarge x limits = 0.2,
 x post scale = 0.75
]

\addplot[fill=LimeGreen  ]
table[
x expr = \coordindex,
y = efficiency	,
]{content/data/dnn_area.txt};

\end{axis}
\end{tikzpicture}
\label{fig:performance_4x4}
}
    \caption{Comparison of a spatial CGRA and \textit{Plaid} (normalized to \textit{Plaid}) on three DNN applications adapted from TinyML~\cite{banbury2021mlperf}.  }
    \label{fig:dnn_comp}
\end{figure}
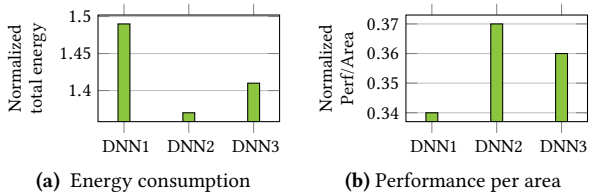

A \textit{2$\times$2} prototype of \textit{Plaid} reveals that the CGRA fabric occupies 33,366 $\mu m^2$ of space, while the scratchpad memories take up an additional 30,000 $\mu m^2$. As illustrated in Figure~\ref{fig:plaid_area}, the breakdown of the CGRA fabric area shows some interesting insights. The communication hardware, including routers and their configuration elements, makes up about 40\% of the area. Meanwhile, the compute hardware and its configuration take up 50\% of the on-chip area. \textit{Plaid} trims down a significant portion of the redundant communication resources and achieves a balance by aligning compute and communication resources, leading to high utilization of both. 

Our evaluation shows that \textit{Plaid} delivers 1.40$\times$ performance and saves 48\% of the area compared to the energy-minimal CGRA while achieving almost the same power. Compared to high-performance spatio-temporal CGRA, \textit{Plaid} reduces 43\% power and saves 46\% area without sacrificing performance and generality. Furthermore, compared to the domain-optimized spatio-temporal CGRA, \textit{Plaid} can still significantly improve energy and area efficiency.  

\subsection{Plaid Achieves High Performance and High  Efficiency}
\textbf{Performance:}
We present the performance of each kernel in terms of the number of cycles it takes for a complete execution on each architecture. Figure~\ref{fig:eval_perf_comp} presents the performance comparison of \textit{Plaid} and the baselines, all normalized to the performance of a spatio-temporal CGRA. All three CGRAs have 16 functional units (ALU and ALSU).

Overall, \textit{Plaid} achieves almost the same performance compared to a spatio-temporal CGRA. While maintaining the same performance, it also achieves a 43\% reduction in power and 46\% reduction area compared to the spatio-temporal CGRA, demonstrating the need to the address the overprovisioning. Moreover, \textit{Plaid} even outperforms the spatio-temporal CGRA on  \textit{durbin\_u4} and \textit{gemver\_u4}. The reason is that the unrolling for these kernels enlarges the DFG but does not significantly complicate the data dependency. For an enlarged mapping space of the problem, the spatio-temporal architecture tries to provide a complex solution to a simple problem and suffers from it. In contrast, \textit{Plaid} can still utilize motifs to generate superior performance. Moreover, the spatio-temporal CGRA achieves better performance on \textit{atax\_u4}, \textit{gramsc\_u2}, and \textit{sedidel\_u2} than \textit{Plaid}. The reason is that more complex and long data dependencies are introduced across multiple parts of the DFG during the unrolling of these kernels, and motifs need more frequent long latency communication with other nodes.
Nevertheless,  the average performance is almost the same between \textit{Plaid} and the spatio-temporal CGRA.

Compared to the baseline spatial CGRA, \textit{Plaid} delivers a 1.40$\times$ improvement in performance. The spatial CGRA needs to partition the DFG to handle complex kernels,  introducing more load and store operations to put intermediate data in SPM.
Spatial CGRA achieves the same performance with \textit{Plaid}  and spatio-temporal CGRA for a few kernels, such as  \textit{fc}, \textit{durbin\_u2}, \textit{dwconv},  \textit{gramsc}, and \textit{jacobi}, The reason is that these kernels have relatively simple data dependency and do not need a lot of additional loads and stores for partitioning.
\newline


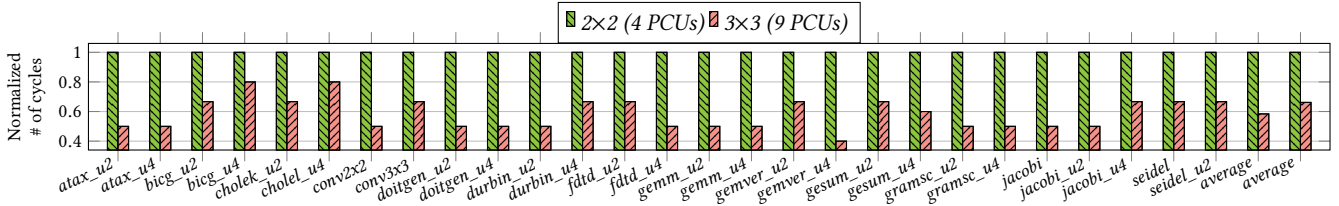
\begin{figure*}
\scriptsize
	\centering
	\begin{tikzpicture}

\pgfplotsset{set layers}
\begin{axis}[
width=21cm, 
height=3cm,
ylabel style={align=center},
ylabel={Normalized \\ \#  of cycles},
bar width=4pt,
ybar=0pt,
xtick=data,
xticklabels from table={content/data/scale.txt}{Benchmark},
x tick label style={rotate=25,anchor=east},
ymajorgrids=true,
yminorgrids=true,
yticklabel={$\pgfmathprintnumber{\tick}$},
legend columns=2,
legend style={at={(0.5,1)}, anchor=south, font=\small},
enlarge x limits = 0.025,
 x post scale = 0.85
]

\addplot[fill=LimeGreen  ,  postaction={pattern=north west lines}]
table[
x expr = \coordindex,
y = plaid2x2	,
]{content/data/scale.txt};
\addlegendentry{\em 2$\times$2 (4 PCUs)}

\addplot[fill=Salmon  ,  postaction={pattern=north east lines}]
table[
x expr = \coordindex,
y = plaid3x3,
]{content/data/scale.txt};
\addlegendentry{\em 3$\times$3 (9 PCUs)}

\end{axis}

\end{tikzpicture}
\hfill
	\caption{Scalability analysis: Performance of 3$\times$3 Plaid compared to 2$\times$2 \textit{Plaid}}
	\label{fig:eval_scalability}
\end{figure*}

\begin{figure*}
    \scriptsize
	\centering
	\begin{tikzpicture}

\pgfplotsset{set layers}
\begin{axis}[
width=21cm, 
height=3cm,
ylabel style={align=center},
ylabel={Normalized \\ \#  of cycles},
bar width=4pt,
ybar=0pt,
xtick=data,
xticklabels from table={content/data/compiler_method_comp.txt}{Benchmark},
x tick label style={rotate=25,anchor=east},
ymajorgrids=true,
yminorgrids=true,
yticklabel={$\pgfmathprintnumber{\tick}$},
legend columns=3,
legend style={at={(0.5,1)}, anchor=south, font=\small},
enlarge x limits = 0.025,
 x post scale = 0.85
]

\addplot[fill= Dandelion ,  postaction={pattern=horizontal lines}]
table[
x expr = \coordindex,
y = pathfinder	,
]{content/data/compiler_method_comp.txt};
\addlegendentry{\em PathFinder}

\addplot[fill= Salmon ,  postaction={pattern=north west lines}]
table[
x expr = \coordindex,
y = sa	,
]{content/data/compiler_method_comp.txt};
\addlegendentry{\em SA}

\addplot[fill=  LimeGreen,  postaction={pattern=north east lines}]
table[
x expr = \coordindex,
y = plaid,
]{content/data/compiler_method_comp.txt};
\addlegendentry{\em Plaid mapper}

\end{axis}

\end{tikzpicture}
	\caption{Performance of \textit{Plaid} on using \textit{Plaid's} mapper compared to using a generic mapper }
	\label{fig:compiler_method_comp}
\end{figure*}
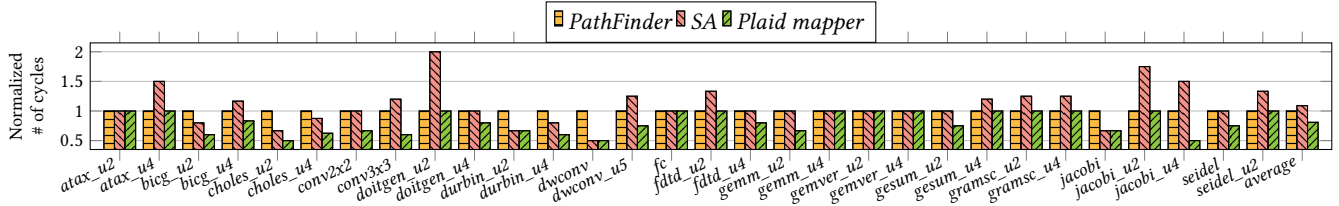



\noindent\textbf{Energy consumption}
Figure~\ref{fig:eval_energy_comp} shows fabric energy consumption comparison normalized to the spatio-temporal CGRA.  \textit{Plaid} achieves 42.0\% and 27.7\% energy reduction compared to spatio-temporal CGRA and spatial CGRA, respectively. The similar performance and    \textit{Plaid's} significant reduction in power directly translates to a reduction in energy consumption over the spatio-temporal CGRA. Furthermore, \textit{Plaid} improves the energy efficiency compared to spatial CGRA, because it can deliver much better performance with roughly the same power.
A typical spatio-temporal or spatial CGRA usually only achieves high performance or high energy efficiency. \textbf{\textit{Plaid} demonstrates that through aligning resource provisioning, a CGRA can be high-performance, general, and efficient at the same time.}
\newline

\noindent\textbf{Performance per area} Figure~\ref{fig:eval_perf_per_area} presents the performance per area comparison normalized to the baseline spatio-temporal CGRA. \textit{Plaid} achieves significant improvement in area efficiency compared to the spatial and spatio-temporal CGRAs. Spatial CGRA can achieve higher energy efficiency but lower area efficiency compared to spatio-temporal CGRA. This is because the spatial CGRAs clock-gate the configuration memory and enable spatial dataflow-based mapping, directly reducing the power, while still requiring similar area. Due to the relatively complex data dependency in \textit{linear algebra},  spatial CGRA achieves less performance per area than the other two domains. \textit{Plaid} achieves a stable improvement compared to spatio-temporal CGRA across the three domains, demonstrating the capability to handle various applications. 
\newline

\noindent\textbf{Application-level comparison} Figure~\ref{fig:dnn_comp} presents the energy consumption and performance per area comparison between \textit{Plaid} and spatial CGRA. We do not include spatio-temporal CGRA as it achieves the same performance with \textit{Plaid} on every machine learning kernel. Spatial CGRA consumes 1.42$\times$ energy and achieves 36\% performance per area compared to \textit{Plaid} on average. With application-level comparison, we can find \textit{Plaid}  still achieves significant improvement in energy consumption and performance per area.

In summary, \textit{Plaid} can achieve high performance, high energy efficiency, and high performance per area, while spatio-temporal CGRA can only achieve high performance, and spatial CGRA can achieve high energy efficiency. Nevertheless, compared to an energy-efficient spatial CGRA, \textit{Plaid} achieves higher energy efficiency. This demonstrates that we optimally align the computing and communication resource provisioning with the proposed design.

\subsection{ Plaid's Hierarchical Execution is Scalable}

Figure~\ref{fig:eval_scalability} shows the performance comparison between 2$\times$2 PCU array and 3$\times$3 PCU array. We exclude DFGs that the 3x3 \textit{Plaid} cannot enhance the performance due to inter-iteration data dependencies. 3$\times$3 \textit{Plaid} can achieve 1.71$\times$ performance compared to 2$\times$2 \textit{Plaid}, demonstrating the scalability of the \textit{Plaid} architecture. The performance increase with the 3x3 \textit{Plaid} does not reach the theoretical maximum for two reasons.  First, the performance might saturate at 2$\times$2 \textit{Plaid},  limiting the additional benefits from the larger PCU array. Secondly, the relationship between the number of nodes in a DFG and the number of functional units affects performance scaling. For instance, if a DFG comprises 40 nodes, the theoretical minimum resource II for the 2x2 \textit{Plaid} (with 16 functional units) and the 3x3 \textit{Plaid} (with 36 functional units) would be 3 and 2, respectively. If we can achieve minimal II on both, 3$\times$3 \textit{Plaid} can only achieve 1.5$\times$ performance.

\begin{figure}
   \scriptsize
	\centering
	\subfloat [ Energy consumption]{
\begin{tikzpicture}
\pgfplotsset{set layers}
\begin{axis}[
width=4.5cm, 
height=3cm,
ylabel style={align=center},
ylabel={Normalized \\ energy},
bar width=3.5pt,
ybar=0pt,
xtick=data,
xticklabels from table={content/data/domain_opt_energy.txt}{Benchmark},
x tick label style={rotate=20,anchor=east},
ymajorgrids=true,
yminorgrids=true,
yticklabel={$\pgfmathprintnumber{\tick}$},
legend columns=2,
legend style={at={(0.8,0.9)}, anchor=south, font=\small},
enlarge x limits = 0.1,
 x post scale = 0.85
]

\addplot[fill=LimeGreen]
table[
x expr = \coordindex,
y = efficiency	,
]{content/data/domain_opt_energy.txt};

\end{axis}
\end{tikzpicture}
\label{fig:performance_3x3}
}
\hfil
\subfloat  [Performance  per area]{
\begin{tikzpicture}

\pgfplotsset{set layers}
\begin{axis}[
width=4.5cm, 
height=3cm,
ylabel style={align=center},
ylabel={Normalized \\ Perf/Area},
bar width=3.5pt,
ybar=0pt,
xtick=data,
xticklabels from table={content/data/domain_opt_area.txt}{Benchmark},
x tick label style={rotate=20,anchor=east},
ymajorgrids=true,
yminorgrids=true,
yticklabel={$\pgfmathprintnumber{\tick}$},
legend columns=2,
legend style={at={(0.8,0.9)}, anchor=south, font=\small},
enlarge x limits = 0.1,
 x post scale = 0.85
]

\addplot[fill=LimeGreen  ]
table[
x expr = \coordindex,
y = efficiency	,
]{content/data/domain_opt_area.txt};

\end{axis}
\end{tikzpicture}
\label{fig:performance_4x4}
}
	\caption{Comparison of domain-specialization in Spatio-Temporal (ST) and \textit{Plaid}. Numbers are normalized to \textit{Plaid}.}
	\label{fig:domain_opt_st_cgra}
\end{figure}
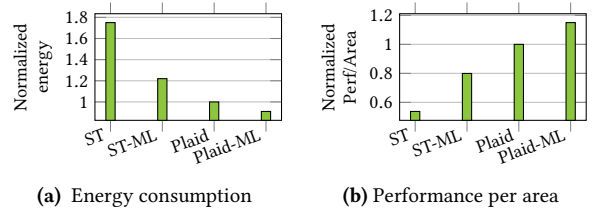

Figure~\ref{fig:compiler_method_comp} shows the performance comparison on \textit{Plaid} CGRA among the aforementioned CGRA mapper PathFinder, Simulated Annealing (SA), and the \textit{Plaid} mapper. The \textit{Plaid} mapper augments SA to support motif-based hierarchical mapping. Thus, the main difference between the two mappers is motif scheduling. The \textit{Plaid} mapper can achieve 1.25$\times$ and 1.28$\times$ performance improvements compared to the PathFinder and SA mapper, respectively. Despite this, both generic mappers can still utilize the hardware designed for collective routing because the routing path is much shorter via local routers. \textit{Plaid}, in conjunction with PathFinder and SA, can achieve comparable performance on several DFGs with spatio-temporal CGRA, demonstrating the effectiveness of the hardware design. However, both generic mappers cannot fully utilize the \textit{Plaid} architecture for more complex DFGs, as they lack the capability to recognize and exploit motifs within DFGs, and are unable to handle the high routing congestion because of the trimmed-down communication circuitry. In contrast, our compiler can automatically identify these motifs and exploit the hierarchical execution paradigm, thereby maximizing the architectural benefits of \textit{Plaid}.
\subsection{Plaid Enhances Domain-Specialization}


Figure~\ref{fig:domain_opt_st_cgra} presents the comparison for various CGRAs: a general spatio-temporal CGRA (ST), a machine learning-optimized spatio-temporal CGRA (ST-ML), general-purpose \textit{Plaid}, and \textit{Plaid} optimized for machine learning (Plaid-ML). ST-ML is generated by pruning the function and bit width to achieve higher energy efficiency and higher performance per area \cite{revamp}.  As mentioned in Section~\ref{sec:domain_spec}, for the Plaid-ML, connections for the motif compute units are hardwired instead of being routed through local routers. We manually check the machine learning DFGs and design Plaid-ML.  Specifically, Plaid-ML uses 2 hardwired \textit{fan-in} motifs, 1 \textit{unicast} motif, and 1 \textit{fan-out} motif for four PCUs, effectively accommodating the motif requirements of our machine learning DFGs.  In terms of performance and energy efficiency, general-purpose \textit{Plaid} outperforms the domain-optimized ST-ML by reducing 18\%  energy consumption and achieving a 1.26 $\times$ performance per area. When comparing Plaid-ML to domain-optimized ST-ML, Plaid-ML shows a significant improvement, reducing energy consumption by 25.5\% and achieving a 1.46 $\times$ performance per area. 


\section{Related Work}

\textbf{Compute and communication resource provisioning:}
HyCUBE~\cite{karunaratne2017hycube} designs a single-cycle multi-hop NoC to improve the capability to handle complex data dependencies. 
Zhang et al. \cite{zhang2019scalable} propose a scalable hybrid network to improve energy efficiency. 
Marionette~\cite{deng2023towards} enhances network efficiency by separating the control and data networks. 
Softbrain~\cite{nowatzki2017stream}, Tartan~\cite{mishra2006tartan}, and Piperench~\cite{goldstein2000piperench} feature dedicated PEs for each instruction. 
Snafu~\cite{gobieski2021snafu}, Manic~\cite{gobieski2019manic}, Riptide~\cite{gobieski2022riptide}, Dyser~\cite{govindaraju2012dyser}, 
and Plasticine~\cite{prabhakar2017plasticine} enhance PEs with limited dataflow semantics. 
While these designs emphasize resource provisioning, they fail to check the collective misalignment between compute and communication provisioning.

\noindent\textbf{Enhancing efficiency:}
REVAMP~\cite{revamp} offers a framework that derives low-power heterogeneous CGRAs from homogeneous ones based on user workloads. Vecpac~\cite{tan2023vecpac} presents a precision-aware CGRA that utilizes multiple 16-bit vector units to support higher data precision.
OpenCGRA~\cite{tan2020opencgra} provides a framework for automatically generating hardwired processing elements for typically linear operation ``chains'' in an application. APEX~\cite{melchert2023apex} and BERET~\cite{gupta2011bundled} automatically design, i.e., hard-wire parts of their circuit for frequent sub-DFGs within a given domain, selecting one block to process each code section. They potentially suffer from under-utilization and over-provisioning of resources~\cite{cgra_retrospective}.
ML-CGRA~\cite{luo2023ml} offers high-level compiler optimizations for machine learning applications on CGRAs.
CCA~\cite{clark2004application,clark2005architecture} is one of the early works to identify and accelerate subgraphs. It features composable rows of functional units to accelerate commonly observed dataflow semantics. However, its network structure offers less flexibility with varying depths, corresponding timing constraints, and configurations to explore depending on the targeted applications.

In contrast to the above architectures, Plaid’s reconfigurable network is designed with a more generalized understanding of (dataflow) graph structures independent of any particular application. This allows for flexible and resource-efficient compositions of functional units, capable of supporting any arbitrary range of patterns. Based on compositions of 3-node motifs within DFGs, Plaid's approach is inherently more general, adaptable, and future-proof, as this structural foundation will remain consistent even as application kernels change.

\section{Conclusions}
We introduce \textit{Plaid}, a novel CGRA architecture and compiler that addresses the misalignment of compute and communication resources. 
\textit{Plaid} reduces power by 43\% and area by 46\% compared to baseline spatio-temporal CGRAs while maintaining high performance and generality. 
\textit{Plaid} offers 1.40$\times$  performance and 48\% area savings compared to energy-efficient spatial CGRAs.
Overall, \textit{Plaid} represents a significant advancement in CGRA design, providing a balanced solution that meets the demands of edge devices while achieving high performance, high energy efficiency, and high generality.

Beyond its immediate benefit on CGRA design, our motif-based execution paradigm offers an alternative to current ad-hoc specialization approaches, where disparate accelerator designs emerge due to the high cost of composition. The fundamental insight of leveraging structural motifs for collective execution could inform the design of other non-CGRA architectures, providing a systematic framework to improve efficiency while maintaining generality. 



\begin{acks}
We would like to thank the anonymous reviewers for their insightful comments.
This research is supported by the National Research Foundation, Singapore under its Competitive Research Program Award NRF-CRP23-2019-0003.
\end{acks}

\bibliographystyle{ACM-Reference-Format}

\bibliography{references}

\end{document}